# Production of a hafnium silicate dielectric layer for use as a gate oxide by solid-state reaction

H. T. Johnson-Steigelman, A.V. Brinck, and P.F. Lyman[*]

*Laboratory for Surface Studies and Department of Physics*

*University of Wisconsin-Milwaukee, Milwaukee, Wisconsin 53211*

**Abstract**

The formation of hafnium silicate films ($HfSi_xO_y$) for use as gate oxides with large dielectric constant by solid state reaction of Hf metal and $SiO_2$ was investigated. Thin, fully reacted silicate films could be formed, and were thermally stable in vacuum to temperatures in excess of 800°C. The interface between a hafnium silicate layer and the silicon substrate was shown to be stable against $SiO_2$ formation. The results suggest a new processing route for production of a gate insulator having low equivalent oxide thickness.

---

[*] Corresponding author (plyman@uwm.edu).

The continued miniaturization of circuit elements in ultra-large scale integrated (ULSI) circuits has required the thickness of the $SiO_2$ layer that forms the gate oxide of field-effect transistors to be decreased.  Although improvements in $SiO_2$ processing have allowed high-quality $SiO_2$ layers of the required thickness to be produced to date, a fundamental thickness limitation looms: for gate oxides thinner than ~12 Å, electron tunneling is expected to provide a current leakage path that is unacceptably high even for high-performance devices, and much too high for increasingly popular low-power applications.  Electron tunneling could be thwarted if a material having a higher dielectric constant than $SiO_2$ could be found that could serve as the gate oxide.  With such a "high-κ" insulator, the gate oxide could be made physically thicker (for a given capacitance), dramatically reducing tunneling.  However, candidate high-κ materials face large impediments to successful integration into the mature Si processing technology, which is based firmly on the near-ideal properties of the Si-$SiO_2$ interface.  For example, the new material must have an electrical interface with few fixed charges or recombination centers, and that does not limit carrier mobility in the transistor channel.  To reduce electrical leakage and dopant diffusion along grain boundaries, the material should readily adopt an amorphous phase.  Moreover, to prevent the advantage gained by employing a high-κ dielectric to be negated, the material must not be prone to reactions with the Si substrate that result in interfacial $SiO_2$ layer formation.

In recent years, an intensive search for such materials has been undertaken.  Familiar high-κ materials used in memory applications such as $Ta_2O_5$ and $SrTiO_3$ were investigated initially, but were shown to undergo deleterious reactions with the Si substrate [1, 2]; this problem is widespread, since the heat of formation of the Si-O bond is sizable. Inspired by a comprehensive examination of the thermodynamic stability of all known binary oxides in contact with Si at 1000 K [3], much recent effort has been invested in compounds of Zr [4 , 5, 6, 7, 8, 9, 10 ] and Hf [7, 11, 12]. These elements are among the most chemically similar found in nature, so the findings of one are often applicable to the other. Oxides of both elements have estimable dielectric constants (~25). However, Hf is slightly more electropositive than Zr, which may provide an advantage over the lighter element. Both elements enjoy what may well turn out to be a significant thermochemical advantage: Hf and Zr can form stable silicate phases $MSi_xO_y$ (M=Hf or Zr), and these compounds appear to be stable in contact with Si at temperatures approaching required processing temperatures [4,5,6,7,9,10,12]. It appears that a relatively small amount of Hf or Zr can significantly increase the dielectric constant [7, 8]. Also, first-principles calculations predict that deleterious Hf-Si (or Zr-Si) bonding will be energetically unfavorable compared to Si-O bonds at the interface [13]. An excellent discussion [7] and a comprehensive review [14] of the pertinent issues have been recently given by Wilk, Wallace, and Anthony. The present work seeks to corroborate this predicted stability for the case of Hf, and to characterize a potential processing route leading to thin layers of hafnium silicate.

Samples were analyzed by x-ray photoelectron spectroscopy (XPS) and low energy electron diffraction (LEED) in a VG Mk. II ESCALab ultrahigh vacuum (UHV) chamber with a base pressure of less than $1\times10^{-10}$ torr. Unmonochromated Al $K_\alpha$ radiation was used to excite the photoelectrons, and spectra were acquired with near-normal electron emission and instrumental resolution of ≤ 0.95 eV. Binding energies were referenced to the bulk Ag $3d_{5/2}$ feature, assumed to lie at 368.2 eV.

Commercial Si(001) samples (*p*-type, 5 Ω-cm) with a native oxide (~15 Å thick) were inserted into the UHV chamber as received, and were degassed by resistive heating to 500°C, rendering a surface nearly free of C contamination. Submonolayer to monolayer (ML) amounts of 99.9% pure Hf (excluding ~3% Zr) were deposited on the substrates using an electrostatic electron beam evaporator [15]. The samples were held at room temperature (RT), and the pressure typically rose to ~5×10$^{-9}$ torr during deposition. After deposition and after each subsequent anneal, XPS and LEED measurements were repeated. To facilitate identification of the spectral features of the UHV-prepared samples, we also prepared and analyzed "bulk" samples of Hf compounds formed by electron-beam evaporation in high vacuum of thicker Hf films (~750 Å) onto Si(001) wafers. These specimens were further processed to create $HfSi_2$, $HfO_2$, and $HfSi_xO_y$ samples [16], which provided convenient reference spectra for the present study.

Upon deposition of 1.1 ML Hf onto the clean native oxide of Si(001), the Hf 4*f* spectrum showed a single broad feature centered at 18.3 eV BE. (Lowest trace in Fig. 1.) Upon annealing to progressively higher temperatures, two trends emerged. The Hf spectral features sharpened, and the centroid of the peak shifted 0.7 eV to deeper BE. For the Hf/SiO$_2$/Si(001) sample annealed to 800°C (top curve in Fig. 1), the 4*f* spectral features were completely dominated by a doublet characteristic of the silicate. (See Fig. 2.) The linewidth of these features is 1.78 eV, which is significantly broader than the spectral resolution (< 0.95 eV). Identical widths were found for "bulk" silicates formed by oxidation of $HfSi_2$. The silicate feature is broader than the corresponding feature in $HfO_2$ spectra (1.42 eV). In contrast, both Hf metal (1.05 eV) and $HfSi_2$ features (0.95 eV) [16] are significantly narrower than either the oxide or silicate. Phonons are likely responsible for part of the increased width of the features of both dielectrics, but the silicate is significantly broader than the oxide line. In an amorphous compound, Hf atoms will be located in a distribution of slightly different local configurations. Thus, the inhomogeneous broadening exhibited in the silicate film indicates that it is amorphous, while the oxide is polycrystalline. Moreover, no ordered LEED patterns were observed for any annealing temperature ≤800°C. Previous studies have shown that hafnium silicates with Hf concentrations in this range readily form in an amorphous phase [17].

For RT deposition, the resulting broad Hf 4$f$ feature (bottom trace in Fig. 1) contains a significant component of a doublet shifted ~1.5 eV to shallower BE. Although it is not possible to uniquely decompose the broad RT feature, it is consistent with one doublet at the silicate BE position (~56% of the intensity) and another doublet (~44% of the intensity) at 1.5 eV shallower BE. (The linewidths, spin-orbit energy splitting, and $4f_{7/2}/4f_{5/2}$ ratio were all constrained to have the values they attain after the 800°C anneal, and a good fit was attained.) We hypothesize that the feature at shallower BE arises from a Hf suboxide formed upon deposition. After annealing to progressively higher temperatures, the suboxide disappears and the 4$f$ features sharpen and shift to deeper BE.

The Si 2$p$ features for this same annealing sequence are shown in Fig. 3. In the lowest spectrum, acquired prior to Hf deposition, the familiar $SiO_2$ feature at 103.5 eV is evident in addition to the bulk Si feature at 99.5 eV. Upon RT Hf deposition, little changes in the Si 2$p$ spectrum. However, upon annealing, the oxide feature shifts continuously to shallower BE. After the 800°C anneal, the oxide feature has shifted by approximately 0.5 eV. In $SiO_2$, each Si atom resides at the center of a tetrahedron, surrounded by four O neighbors. This is also true in the case of ideal hafnium silicates, explaining the rough similarity of the Si BE position for $SiO_2$ and hafnium silicates. The difference is that these O atoms are also bonded to Hf atoms, not other Si atoms. Hf is more electropositive than Si, so insertion of a Hf ion to form a Hf-O-Si complex would reduce the ionicity in the Si-O bond (compared to one in a Si-O-Si complex) [18, 19]. This shift will reduce the Si 2$p$ BE, consistent with our observations. This shift provides evidence that the hafnium silicate/Si interface will be stable against interfacial $SiO_2$ formation (as predicted by Hubbard and Schlom [3]). The thermodynamic driving force for interfacial $SiO_2$ formation when most oxides are placed in contact with Si is the large heat of formation of the $SiO_2$ phase. While Si is rather electropositive, Hf is even more electropositive, and $HfO_2$ has a higher heat of formation than does $SiO_2$. The shift of the Si oxide XPS feature to shallower BE indicates [18, 19] that Hf donates charge to the $SiO_2$ complexes in the newly formed silicate compound. This shift, therefore, corroborates that Hf is able to reduce $SiO_2$; conversely, Si will be unable to reduce $HfO_2$, and interfacial $SiO_2$ formation will be thermodynamically unfavorable. These results are not surprising, since there have been previous reports that, for Hf and Zr oxides and silicates in contact with Si, an interfacial $SiO_2$ layer is not evident by transmission electron microscopy (TEM) [4, 5, 6, 7, 9, 10, 12]. However, to our knowledge, there have been no previous observations of charge transfer during silicate formation to corroborate those structural findings.

Another key issue for device applications is the thermal stability of hafnium silicate films. We have found the thermal stability to vary with film thickness [16]. The likely mechanism for compositional breakdown of these films in a reducing atmosphere (such as vacuum) is O loss through SiO desorption. For this process to take place, excess Si must diffuse through the film to the vacuum interface. For hafnium silicate films formed by oxidizing $HfSi_2$, films of ~200 Å started to break down upon extended annealing to 850°C, while very thick silicate films were stable upon annealing to 1020°C [16]. In the present study, the very thin silicate layers formed by solid state reaction were observed to break down to form $HfSi_2$ upon annealing to 970°C. (Temperatures between 800° and 970°C were not tested.) In any event, our results indicate that there is a thermal window within which silicate films can be formed by solid state reaction, but will not decompose under vacuum annealing. At present, it is unclear whether sufficiently thin films will be stable at the processing temperatures and atmospheric conditions required for ULSI manufacturing.

Solid phase reaction possesses advantages compared to direct deposition of the silicate, as used in the studies by Wilk et al. [4, 7, 11]. The most significant potential advantage is that it would allow the well-developed procedures for producing a high-quality $SiO_2$ layer with a good Si interface to be retained in the processing. In contrast, the direct deposition of a silicate layer onto bare Si would require that any $SiO_2$ layer be first removed [7], exposing the bare surface to the ambient or to the poor vacuum found in sputter deposition tools. Solid-state reaction therefore promises better and less contaminated dielectric-Si interfaces. Furthermore, it appears that high-quality silicates can not be deposited directly using electron-beam evaporation. Although Wilk et al. performed direct deposition using both sputtering and electron-beam deposition, only sputter deposition produced films with acceptable electrical characteristics, and that lacked deleterious Hf-Si bonding. The present method allows electron beam deposition to be used, so the structure being fabricated would not suffer from the ion damage attendant with sputter deposition. Also, it is likely our method can be extended to allow deposition of the required Hf metal by chemical vapor deposition (CVD), rather than electron –beam evaporation. Growth of metallic Hf by CVD will likely present far fewer process problems (such as C incorporation) than attempts to grow stoichiometric silicate layers using that technique.

Clearly, films produced by solid-phase reaction of Hf metal and $SiO_2$ will be O-deficient. The reacted sample could be fully oxidized in, e.g., a rapid thermal annealing (RTA) step. Although a re-oxidation would raise the possibility of interfacial $SiO_2$ formation, O diffusion through hafnium silicates has been reported to be relatively slow [20].

In conclusion, we have investigated the formation of hafnium silicate films by solid state reaction of Hf metal and $SiO_2$. We find that fully reacted, amorphous films of $HfSi_xO_y$ may be formed by this process, and they are thermally stable in vacuum to temperatures in excess of 800°C. We present compelling evidence that the silicate-substrate interface is stable against formation of a $SiO_2$ layer, a key requirement for producing a gate oxide stack with high capacitance.

**Acknowledgements**

This work supported by the National Science Foundation under contract number DMR-9984442-01 and by an REU site grant.

1  G.B. Alers, D.J. Werder, Y. Chabal, H.C. Lu, E.P. Gusev, E. Garfunkel, T. Gustafsson, and R.S. Urdahl, Appl. Phys. Lett. **73**, 1517 (1998).

2  K. Eisenbeiser, J.M. Finder, Z. Yu, J. Ramdani, J.A. Curless, J.A. Hallmark, R. Droopad, W.J. Ooms, L. Salem, S. Bradshaw, and C.D. Overgaard, Appl. Phys. Lett. **76**, 1324 (2000).

3 K.J. Hubbard and D.G. Schlom, J. Mater. Res. **11**, 2757 (1996).

4  G.D. Wilk and R.M. Wallace, Appl. Phys. Lett. **76**, 112 (2000).

5  M. Copel, M. Gribelyuk, and E. Gusev, Appl. Phys. Lett. **76**, 436 (2000).

6  W.-J. Qi, R. Nieh, E. Dharmarajan, B.H. Lee, Y. Jeon, L. Kang, K. Onishi, and J.C. Lee, Appl. Phys. Lett. **77**, 1704 (2000).

7  G.D. Wilk, R.M. Wallace, and J.M. Anthony, J. Appl. Phys. **87**, 484 (2000).

8 G. Lucovsky and G.B. Rayner, Jr., Appl. Phys. Lett. **77**, 2912 (2000).

9 T.S. Jeon, J.M. White, and D.L. Kwong, Appl. Phys. Lett. **78**, 368 (2001).

10 J. Morais, E.B.O. da Rosa, L. Miotti, R.P. Pezzi, I.J.R. Baumvol, A.L.P. Rotondaro, M.J. Bevan, and L. Colombo, Appl. Phys. Lett. **78**, 2446 (2001).

11 G.D. Wilk and R.M. Wallace, Appl. Phys. Lett. **74**, 2854 (1999).

12 B.H. Lee, L. Kang, R. Nieh, W.-J. Qi, and J.C. Lee, Appl. Phys. Lett. **76**, 1926 (2000).

13  S. Kawamoto, J. Jameson, P. Griffin, K. Cho, and R. Dutton, IEEE Electron. Dev. Lett. **22**, 14 (2001).

14  G.D. Wilk, R.M. Wallace, and J.M. Anthony, J. Appl. Phys. **89**, 5243 (2001).

15 B.T. Jonker, J. Vac. Sci. Technol. A **8**, 3883 (1990).

16  A.V. Brinck, H.T. Johnson-Steigelman, and P.F. Lyman, to be published.

17  M.A. Russack, C.V. Jahnes, and E.P. Katz, J. Vac. Sci. Technol. A **7**, 1248 (1989).

18 T.L. Barr, Crit. Rev. Anal. Chem. **22**, 115 (1991).

19  T.L. Barr, J. Vac. Sci. Technol. A **9**, 1793 (1991).

20 S.P. Murarka and C.C. Chang, Appl. Phys. Lett. **37**, 639 (1980).

**Figure Captions**

Fig. 1. Hf 4*f* XPS data for 1.1 ML Hf/15 Å SiO$_2$/Si(001) films annealed to the indicated temperatures. The feature shifts to deeper BE upon annealing and sharpens as suboxide component is eliminated.

Fig. 2. Hf 4*f* XPS data and fit for sample annealed to 800°C. A Shirley background has been subtracted. The data may be explained by a single doublet characteristic of hafnium silicate.

Fig. 3 Si 2*p* XPS data for 1.1 ML Hf/15 Å SiO$_2$/Si(001) films annealed to the indicated temperatures. The feature associated with silicon oxide at 103.5 eV shifts to shallower BE upon annealing, indicating charge transfer to the Si-O complexes. The charge transfer implies that the silicate/silicon interface will be stable against SiO$_2$ formation.

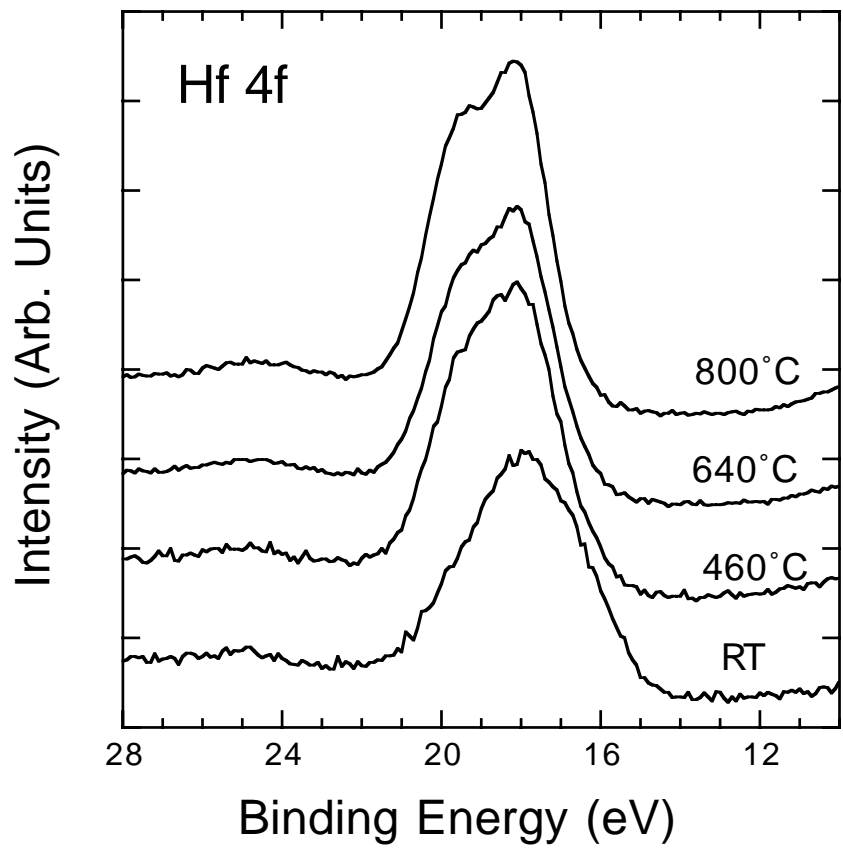

Fig.1 Johnson-Steigelman et al.

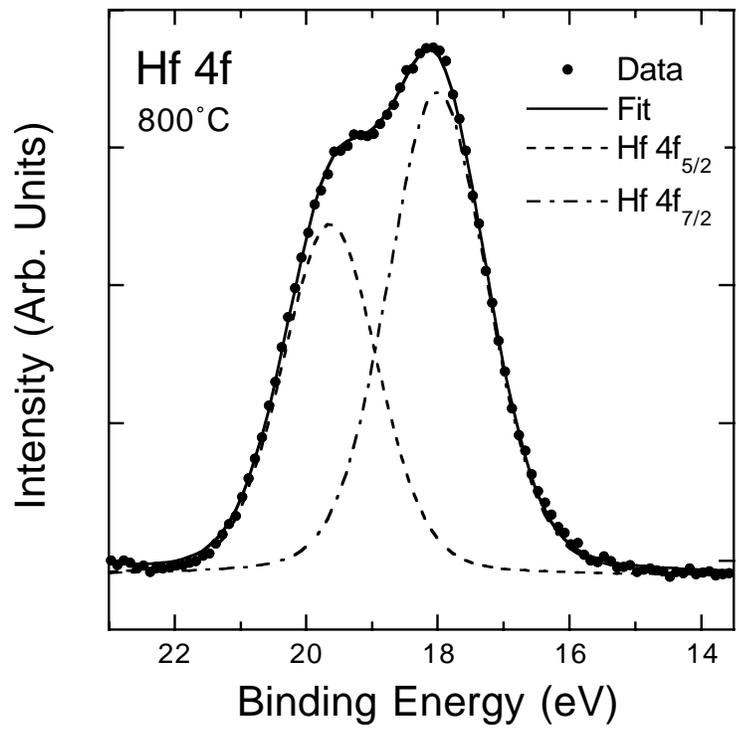

Fig.2 Johnson-Steigelman et al.

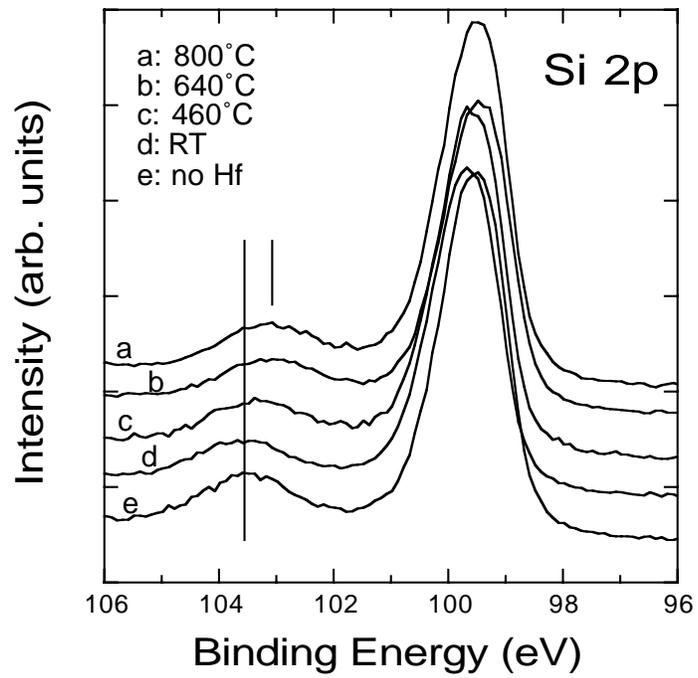

Fig.3 Johnson-Steigelman et al.